\documentstyle[aps,twocolumn]{revtex}

\begin{document}
\author{Horacio M. Pastawski, Patricia R. Levstein and Gonzalo Usaj}
\address{Facultad de Matem\'{a}tica, Astronom\'{\i}a y F\'{\i}sica.
Universidad\\
Nacional de C\'{o}rdoba. \\
Ciudad Universitaria. 5000 C\'ordoba-ARGENTINA}
\title{Quantum Dynamical Echoes in the Spin 'Diffusion' in Mesoscopic Systems. }
\date{received 31 May 1995 }
\maketitle

\begin{abstract}
The evolution of local spin polarization in finite systems involves
interference phenomena that give rise to {\bf quantum dynamical echoes }and
non-ergodic behavior. We predict the conditions to observe these echoes by
exploiting the NMR sequences devised by Zhang et al. [Phys. Rev. Lett. {\bf %
69}, 2149 (1992)], which uses a rare $^{13}$C as {\bf local probe }for a
dipolar coupled $^1$H spin system. The non-ideality of this probe when
testing mesoscopic systems is carefully analyzed revealing the origin of
various striking experimental features.
\end{abstract}

In recent years interest has aroused on transport properties of mesoscopic
systems \cite{Lee}. Examples range from photons propagating through a
colloidal solution to electrons in nanostructures. In most of the systems
the quantum phenomena manifest as transport properties which are not linear
with the size of the system. Since the Boltzmann equation does not retain
interferences after collisions, it is necessary the use of the
Schr\"{o}dinger equation (e.g. in Liouville \cite{Liouville} or Keldysh's
form \cite{Pastawski}). Thus, a localized excitation, after diffusing away
and reaching the boundaries, might return as {\bf quantum} {\bf dynamical
echoes }\cite{Prigodin}{\bf \ } resembling the familiar sound echoes. A
trivial case occurs when a particle is placed in one side of a symmetrical
double well potential and oscillates among them \cite{Levstein}. While in
these examples the high group velocities of the excitations restrict the
observation of interference phenomena in the time domain, the local
excitations in a system of nuclear spins with magnetic dipolar interactions
evolve with characteristic times on the hundreds of micro-seconds and can be
tested by using an NMR technique devised recently by Zhang, Meier and Ernst 
\cite{ZME} (ZME). While they addressed a realization of the Lochsmidt daemon 
\cite{Rhim}, they showed that, by applying an appropriate radio-frequency
pulse sequence involving cross-polarization \cite{HH}, the spin $S$ of a
rare $^{13}$C$\,\,\,\,$(1.1\% abundance) bonded to a $^1$H\thinspace $\,\,$%
with spin $I_1$ can be used as a {\bf local probe} that injects
magnetization in the proton and captures it at a later time. The set of
magnetic dipolar coupled $^1$H within a molecule constitutes the mesoscopic
system where the spin dynamics can be monitored. The rest of the crystal
constitutes a weakly interacting thermal bath.

The experimental \cite{ZME} time evolution of the polarization of a proton
spin in a crystal of ferrocene, (C$_5$H$_5$)$_2$Fe$,$ shows a nearly
exponential decay of the local magnetization toward an asymptotic value of $%
0.2$ of the initial polarization. This was interpreted as an ''unhindered''
spin diffusion among the five protons of a single ring. This is based on the
assumption that the $^{13}$C is an {\bf ideal local probe}, i.e. the
polarization transfer between the $^1$H and $^{13}$C is complete and it
takes a short time as compared with the characteristic time of the $^1$H-$^1$%
H interactions \cite{Muller}. However, there are two aspects in the results
of this ingenious experiment which do not fit with our current understanding
of mesoscopic quantum dynamics: One is the linear decay of the magnetization
observed for short times which is obviously inconsistent with the expected
quadratic decay. The other is the monothonic decay which does not reveal any
of the interference features characteristic of quantum evolution in systems
with few degrees of freedom. The search for the conditions to actually
observe quantum dynamical echoe and answer to these questions lead us to
solve the Schr\"{o}dinger equation for the model system proposed by ZME.

Let us consider an ideal situation in which {\it there is} a perfect local
probe which is able to inject magnetization in a single $^1$H nucleus and to
measure it later. The system consists of $N$ $I$-spins governed by the
magnetic dipolar ($\alpha =1$) Hamiltonian truncated with respect to the
dominant Zeeman interaction:

\begin{equation}
{\cal H}_{II}=\sum_{j>k}\sum_k^{\,\,}d_{jk}\,\left[ \alpha 2I_j^zI_k^z-\frac 
12\left( I_j^{+}I_k^{-}+I_j^{-}I_k^{+}\,\right) \right] .  \label{Hii}
\end{equation}
Site $j$ spin operators $I_j$ are dimensionless. Subscripts start
sequentially from $j=1.$ The interaction parameters are 
\begin{equation}
d_{jk}^{}=-\frac{\mu _{0\,\,}\gamma _I^2\,\,\,\,\hbar ^2}{4\pi r_{jk}^3}\,\,%
\frac 12\left\langle 3\cos ^2\theta _{jk}-1\right\rangle ,  \label{d}
\end{equation}
where $\gamma _I^{}$ is the gyromagnetic factor of the $I$-spins, $r_{jk}$
are internuclear distances and $\theta _{jk\text{ }}$ their angles with the
static magnetic field. A simplifying fact about ferrocene is that the ring
performs fast rotations around its five-fold symmetry axis with a very short
correlation time\cite{Seiler} ($\approx 10^{-12}{\rm s}$) and therefore the
interaction parameter is time averaged. The new constants depend \cite
{Slichter} only on the angle $\theta $ between the molecular axis and the
magnetic field and the angles $\gamma _{j,k}$ between the internuclear
vectors and the rotating axis. The factor in brackets in Eq.(\ref{d})
becomes $\left\langle \frac 12(3\cos ^2\theta -1)(3\cos ^2\gamma
_{j,k}-1)\right\rangle =1/2$ for a magnetic field normal to the axis. All
sites in a molecule are then equivalent and the five fold symmetry is
recovered, giving $d_1=1576Hz\times 2\pi \hbar $ for nearest neighbors. A
locally polarized initial state $\left| i\right\rangle $ evolves toward a
final state $\left| f\right\rangle $ with probability$:$

\begin{equation}
P_{fi}(t_1)=\left| \langle f\mid \exp [-\frac{{\rm i}}\hbar {\cal H}%
_{II}{}t_1]\mid i\rangle \right| ^2.  \label{Pt}
\end{equation}
In the high temperature limit all the $N_i\equiv N_f$ configurations
compatible with an up projection of the spin $I_1$ are equally probable.
Then the magnetization will be:

\begin{equation}
M(t)=2\left[ \sum_f^{N_f}\sum_i^{N_i}\frac 1{N_i}P_{fi}(t_1)-\frac 12\right]
.  \label{Mt}
\end{equation}
For an {\it ideal} system of five spins arranged in a ring configuration,
there is $1$ configuration with total projection $5/2$, and there are $5$, $%
10$, $10$, $5$ and $1$ configurations with projections $3/2$, $1/2$, $-1/2$, 
$-3/2$ and $-5/2$ respectively. On each of these subspaces there are $1$, $4$%
, $6$, $4$, $1$ and $0$ possible initial and final states with spin $I_1$
polarized summing up $N_i=16$. The evolution under Hamiltonian (\ref{Hii})
is called 'diffusion' in the {\it laboratory frame}. For the results of
local polarization presented in Fig. 1 we multiply by $-[\frac 12]$ the
above interaction constants, which correspond to the experimental case of
'diffusion' in the {\it rotating frame}. The dashed line represents the
evolution when only nearest neighbor interactions are taken into account,
while the full line considers all the interactions. At {\it intermediate
times}, the first shows a clear quantum dynamical echo at around $1400\mu 
{\rm s}\approx 5\times \sqrt{2}\times \hbar /([\frac 12]d_1)$, which can be
interpreted as the time the excitation needs to wind around a ring of length 
$L=5a$ at an average speed $v_M/\sqrt{2}$, with maximum group velocity $%
v_M=a[\frac 12]d_1/\hbar $. This is confirmed by the study of correlation
functions at different sites: the polarization at site $2$ and $5$ has the
first maximum at $280\mu {\rm s}$ and at sites $3$ and $4$ around $560\mu 
{\rm s}$. Including second neighbors interactions, maxima are split and
weakened, this can be assigned to the effect of short-cutting the path
around. Although dynamical echoes are still visible, one may conclude that
they are weaker when more degrees of freedom become effectively mixed. In
the {\it long time} regime a striking effect is noticeable in the inset of
Fig. 1: local polarization does not distribute around the value $1/5$ but
rather oscillates around $\overline{M}$=$\,\lim_{t\rightarrow \infty
}\,t^{-1}\int_0^tM(t^{\prime }){\rm d}t^{\prime }\approx 0.32.$ This
non-ergodic behavior is a consequence of the degeneracy of opposite
quasi-momenta in perfect rings. We have observed this for single rings as
well as for pairs of coaxial rings with $N\leq 10$ spins interacting
according to dipolar ($\alpha =1$), Ising like ($\alpha \gg 1$), Heisenberg (%
$\alpha =-\frac 12$) and XY ($\alpha =0$) interactions by studying $%
\overline{M}.$ In the last situation we obtained the analytical high
temperature solution of Eq. ($\ref{Mt})$, which for odd rings gives: $%
\overline{M}_{{\rm i}}=1/N+(N-1)/N^2,$ where the second term accounts for
the degenerate quasi-momentum states. The {\it complete ferrocene molecule}
gives $\overline{M}_{{\rm i}}\approx 0.2.$ Here we distinguish rings
rotating independenly or as a whole, keeping an eclipsed or staggered
conformation. While the first present a weaker version of the dynamical
echoes of the single ring, in the other two the echoes and valleys merge in
a sort of plateau for intermediate times. For very {\it short times} the
magnetization decreases quadratically with $t_1^{}.$ The coefficient is the
configuration average of the sum of squares of the coupling constants
effective to the flip-flop process in each configuration$.$

These results clearly do not represent the experimental data in ZME \cite
{ZME}. Although quantum dynamical echoes are indeed attenuated by including
the interactions with other rings and molecules, the linear decay of the
magnetization suggests that some quantum evolution has already occurred for
the experimental time $t_1=0$. Therefore, it is important to consider that
magnetization is injected and monitored using the spin $S$, which
constitutes a {\bf non-ideal probe}. The complete pulse sequence is
schematized in Fig. 2. The most relevant part is as follows: {\bf A) }The
magnetization from an initially $y$ polarized $S$ spin is transferred to the 
$y$-axis of the $I_1^{}$ spin (and in lesser degree to its neighbors)
through a cross polarization pulse of duration $t_d$, when both the abundant
and rare spins are irradiated at their respective resonant frequencies with
field strengths fulfilling the Hartmann-Hahn \cite{HH} condition $\gamma
_IB_{1I}^y=\gamma _SB_{1S}^y.$ The time $t_d$ is the shortest of those that
maximize the polarization transfer \cite{Muller}. {\bf B) }The $I$-spins
evolve in presence of $B_{1I}^y$ during a time $t_1$. Thus, the relevant
evolution of the spin-diffusion sequence occurs in a rotating frame.$\,\,$%
{\bf C)} Another cross-polarization pulse of length $t_p=t_d$ is applied to
transfer back the polarization to the $x$-axis of $S$. {\bf D) }The $S$
polarization is detected while the $I$-system is kept irradiated
(high-resolution condition). Therefore a better simulation of the experiment
includes the $I$'s spins system, the non ideal probe $S$ and their
interaction. The $I$-$S$ dipolar interaction extends the sum in (\ref{Hii})
to the $S$ spin. The interaction coefficient is $b_k^{}$ defined as ($\ref{d}%
)$ with a $\gamma _I$ replaced by a $\gamma _S$ and distances and angles
modified accordingly. Given the difference in gyromagnetic factors the
dipolar $I$-$S$ interaction is truncated dropping terms that mix different
Zeeman subspaces, hence: 
\begin{equation}
{\cal H}_{IS}=\sum_kb_k\,\,2I_k^zS_{}^z,  \label{His}
\end{equation}

When two r.f. fields are applied to both $I$ and $S$ spins, these are
quantized in their corresponding rotating frames and the total Hamiltonian
is:

\begin{equation}
{\cal H}_{}^{y(x)}={\cal H}_{II}^{}+{\cal H}_I^{}+{\cal H}_S^{y(x)}+{\cal H}%
_{IS}^{}  \label{Hyx}
\end{equation}
the superscript indicates the direction of $B_{1S}^{}$.

The {\bf isolated I-spins system }is described by the first two terms, where 
${\cal H}_I=-\hbar \omega _{1I}\sum_kI_k^y$ with $\hbar \omega _{1I}=\gamma
_IB_{1I}^y$ is the effect of the r.f.. In the case of strong irradiation, $%
\mid \hbar \omega _{1I}\mid \gg \mid d_{jk}\mid $ , we can approximate (\ref
{Hii}) in a frame tilted to the rotating $y$ axis by 
\begin{equation}
{\cal H}_{II}^{^{\prime }}=-[%
{\textstyle {1 \over 2}}
]\sum_{j>k}\sum_k^{\,\,}d_{jk}\,\left[ 2I_j^yI_k^y-\frac 12\left(
I_j^{+}I_k^{-}+I_j^{-}I_k^{+}\,\right) \right] .  \label{Hii'}
\end{equation}
As usual, we {\it neglected} non-secular terms of the form $\left(
I_j^{+}I_k^{+}+I_j^{-}I_k^{-}\,\right) .$ It is the factor $-[\frac 12]$
relating (\ref{Hii}) and (\ref{Hii'}) what allows the realization of the
Lochsmidt daemon \cite{Rhim} by experimentally switching the evolution from
one situation to the other which leads to a {\it polarization echo} \cite
{ZME}, not to be confused with the quantum dynamical echoes described above.
The {\bf isolated probe} in the presence of a r.f. field along $y$ $(x)$
direction is described by ${\cal H}_S^{y(x)}=-\hbar \omega
_{1S}S_{}^{\,\,y(x)}$ with $\hbar \omega _{1S}=\gamma _SB_{1S}^{y(x)}$. The 
{\bf system-probe interaction} with an additional truncation in the doubly
rotating frame is:

\begin{equation}
{\cal H}_{IS}^{}\approx \sum_kb_k\frac{({\rm i})}2\,[I_k^{+}S_{}^{-}+(-1)\,%
\,\,I_k^{-}S_{}^{+}],  \label{His'}
\end{equation}
where factors in parenthesis correspond to $B_{1S}^x$. In Fig. 2, the
evolution starting with the polarized $S$ spin involves two Hamiltonians,
during the times $t_d$ and $t_p$ the evolution is given by the total
Hamiltonian ${\cal H}_{}^{y(x)},$ but during $t_1$ as well as during the
acquisition, the irradiation makes the action of ${\cal H}_{IS}^{}$
negligible (high resolution condition) and evolves with ${\cal H}%
_{II}^{\prime }+{\cal H}_I^{}$. The probability of having an initial state $%
\left| i_y\right\rangle $ with the probe polarized along $y$ direction and a
final state $\left| f_x\right\rangle $ with the probe polarized along $x$
is: 
\begin{equation}
P_{f,i}(t_1)=\mid \langle f_x\mid \exp [-%
{\textstyle {{\rm i} \over \hbar}}
{\cal H}^xt_p]\exp [-%
{\textstyle {{\rm i} \over \hbar}}
({\cal H}_I+{\cal H}_{II}^{\prime })t_1]\exp [-%
{\textstyle {{\rm i} \over \hbar}}
{\cal H}^yt_d]\mid i_y\rangle \mid ^2
\end{equation}

The resulting magnetization is shown in Fig. 3 with open circles. For short $%
t_1$ the evolution is linear with time and the dynamical echoes are still
noticeable but appear shifted to shorter times. Both effects are consequence
of the evolution of the magnetization in the $I$-$I$ system during the two
cross polarization periods: by comparing with Fig.1, it is clear that the
overall effect of them is to shift the curve in a time shorter than their
sum. This is verified by superposing the ideal evolution from Fig. 1,
shifted in $t_{{\rm shift}}=126\mu s<(t_d+t_p)$ and renormalized with a
factor of $a=1.22$ obtained from the short time behavior. This explains why
the asymptotic value of the magnetization in the complete sequence has been
raised to $\overline{M}_{{\rm n.i}.}=a\overline{M}_{{\rm i}}\approx 0.4$
from the $\overline{M}_{{\rm i}}\approx 0.32$ around which the ideal
polarization of Fig.1 oscillates. Then $\overline{M}_{{\rm n.i}.}$
overestimates the fraction of magnetization remaining in the proton. The
fact that Figs. 1 and 3 can be superposed indicates that $S$-$I$
interactions do not break the five-fold symmetry appreciably within the
relevant experimental times, this is a consequence of the high resolution
condition.

These results indicate that the experimental \cite{ZME} asymptotic value $%
\overline{M}_{\exp }=0.2=\overline{M}_{{\rm n.i}.}$ originates on the
evolution that has already occurred at $t_1=0,$when polarization starts to
be recorded and is normalized. A more proper normalization would give an
ideal asymptotic value $\overline{M}_{{\rm i}}$ $\approx \overline{M}_{\exp
}/a=$ $0.16,$ lower than $\overline{M}_{{\rm i}}$ for the complete molecule
(two coupled rings). Since, as discussed above, the quantum evolution {\bf %
does not give an equi-distribution of polarization,} to achieve the
experimental $0.2$, the evolution must involve a system with $N\gg 10.$ With
the participation of an important number of spins other than those belonging
to one molecule, the boundary of the system becomes fuzzy, weakening the
interferences that originate the quantum dynamical echoes. Then, it becomes
clear why these do not show up in the reported experiments.

An interesting phenomenon shown up by the non-ideal evolution in Fig 3 is
the small high frequency ($\approx \omega _{1I}$) oscillation around the
superposed ideal curve. This effect is due to an incomplete $S$-$I$ transfer
which allows interference of phases accumulated during evolution $t_1$ by
sub-spaces with different total spin $\sum I_k^{\,\,\,y}$ projection; i.e.
the interference of the dashed and dotted paths of polarization amplitudes
in Fig. 2. They may originate the small 'dispersion' observed in the top of
the polarization echo in the experimental curve.

In summary, we have solved the relevant part of a non-ideal spin 'diffusion'
experiment for a simple five-spins system resorting to the usual assumptions
to generate the system, probe and system-probe Hamiltonians: truncation of
the dipolar Hamiltonian and total decoupling under high resolution
conditions (during the time $t_1$)$.$ We have shown that while $^{13}$C
behaves as a {\bf non-ideal local probe}, its effect can be reasonably
quantified. As occurs in all mesoscopic physics the probes are actually part
of the system \cite{Buttiker} and therefore have a relevant role in the
interpretation of the results. This undesired effect can be experimentally
overcome in part, by studying the diffusion in the laboratory frame which
tends to compensate the non-ideality of the probe and shrinks in a factor $%
\left[ \frac 12\right] $ the time scale, 'cleaning' the evolution from
interactions different than the {\it secular} dipolar one. We predict that a
spin 'diffusion' experiment in molecular crystals containing few interacting
spins per molecule should present {\bf quantum dynamical echoes}. Possible
candidates are molecules with single rings of the form (C$_n^{}$H$_n$)$.$
Molecules with linear topology in the interaction network present less
important second neighbors interaction, making the interference effects even
stronger.

This work was performed at LANAIS de RMN (UNC-CONICET) with financial
support from Fundaci\'on Antorchas, CONICOR and SeCyT-UNC.

\newpage\ 

{\bf Figure 1: } Ideal evolution of a $^1$H-spin polarization as a function
of $t_1$ for a system of $5$ spins in a ring. Dashed line considers only
nearest neighbor interaction $d_1$. Full line is the exact dipolar
evolution. The inset shows fluctuations around the value $\overline{M}_{{\rm %
i}}=0.32$.

\bigskip\ 

{\bf Figure 2:} Pulse sequence for measurement of proton spin 'diffusion' in
the rotating frame. A $\pi /2$ pulse on the abundant $^1$H spins system
creates a polarization that is transferred during $t_C$ to a rare $^{13}$C.
After the decay of the proton spin coherence during $t_S$: ({\bf A}) the $%
^{13}$C polarization is locally injected to the bonded proton, ({\bf B})
'diffusion' in the protons system is allowed and ({\bf C}) it is captured
again in the $^{13}$C where ({\bf D}) it is recorded.

\bigskip\ 

{\bf Figure 3:} Non-ideal evolution of a $^1$H spin polarization in a ring
of ferrocene (open circles) as detected in a strongly coupled $^{13}$C spin.
Calculation uses the pulse sequence in Fig. 2 with $t_d=t_p=85\mu {\rm s}$
with the molecular symmetry axis at an angle of $\pi /2$ with the external
magnetic field. Full line is the ideal evolution in Fig.1, shifted in $%
126\mu {\rm s}$ and normalized accordingly.

\end{document}